\documentclass[12pt]{article}
\usepackage{amsmath}
\usepackage{cite}
\usepackage{graphicx}
\usepackage{subfigure}
\usepackage{times}
\usepackage{multicol}
\usepackage{fullpage}
\numberwithin{equation}{section}

\begin{document}
\pagestyle{empty}

\begin{center}
{\LARGE \bf  The Thermodynamics of a 5D Gravity-Dilaton-Tachyon Solution}

\vspace{1.0cm}

{\sc Thomas M. Kelley}\footnote{E-mail:  kelley@physics.umn.edu}\\
\vspace{.5cm}
{\it\small {School of Physics and Astronomy, University of Minnesota,\\
Minneapolis, MN 55455, USA}}\\

\vspace{.3cm}
\today
\end{center}

\vspace{1cm}
\begin{abstract}
We propose a finite-temperature holographic model with a soft-wall geometry that incorporates two 
scalar fields, dual to the gluon and chiral operators. A series solution is presented as the dynamical, black-hole 
solution to Einstein's equations. We use the solution to calculate the thermal properties of the corresponding 4D 
gauge theory. Gluon and chiral thermal condensates contribute leading-order terms that affect the speed of sound through and entropy of the 4D thermal medium. At a temperature $T_c\sim 900$~MeV, we find a phase 
transition, which is much higher than lattice QCD calculations. However, the transition only exists with nonzero thermal condensates. 
\end{abstract}

\vfill
\begin{flushleft}
\end{flushleft}
\eject
\pagestyle{empty}
\setcounter{page}{1}
\setcounter{footnote}{0}
\pagestyle{plain}

\section{Introduction}

The anti-de Sitter/conformal field theory (AdS/CFT) correspondence~\cite{Maldacena:1997re,Gubser:1998bc,Witten:1998qj} has provided new insight into the dynamics of strongly coupled, four-dimensional (4D) gauge theories from the remarkable duality with a weakly coupled, five-dimensional (5D) gravitational description. Quantum chromodynamics (QCD), the theory of strong interactions, is a gauge theory that becomes strongly coupled in the infrared (IR); therefore, a 5D gravity dual provides a new framework to study nonperturbative phenomena in QCD. This is done in AdS/QCD models, where the conformal symmetry is broken at a particular energy scale by an IR brane (hard-wall~\cite{DaRold:2005zs,Erlich:2005qh}) or gradually via a background scalar field (soft-wall~\cite{Karch:2006pv}). The hard-wall models provide a reasonable fit to the low energy observables except that the mass spectrum
$m_n\sim n$. A much better fit to the mass spectrum occurs in soft-wall models that can mimic the observed Regge trajectories of the QCD hadrons~\cite{Karch:2006pv,Colangelo:2008us}. While the exact 5D dual of QCD remains elusive, further improvements such as chiral symmetry breaking have been incorporated into AdS/QCD models~\cite{Evans:2006ea,Evans:2006dj,Csaki:2006ji,Kwee:2007nq,Gherghetta:2009ac} but still describe only part of the rich
structure of QCD.

Furthermore, there have been attempts to study strongly coupled gauge theories at finite temperature. This requires extending the metric to an AdS-Schwarzschild solution, describing an extra-dimensional generalization of a black hole. The thermodynamic description of black holes is well established~\cite{Bekenstein:1973,Bekenstein:1974,Hawking:1974sw}. Using black-hole solutions, finite-temperature properties of strongly coupled gauge theories can be studied, such as those associated with the deconfinement phase transition and the quark-gluon plasma. Interestingly, experimental evidence confirms that the quark-gluon plasma (QGP) is strongly coupled \cite{Adams:2005dq, Adcox:2004mh, Arsene:2004fa, Back:2004je}, lending support to the theoretical picture obtained via a 5D dual gravity description.

The simplicity of the soft-wall models in providing a reasonable fit to the mass spectrum, together with 
the success of the black-hole solutions at finite temperature, suggests that a simple model combining 
both features can be obtained. In this paper, we provide a finite-temperature description
of a strongly coupled gauge theory using a soft-wall geometry based on the model in  \cite{Batell:2008zm}.
A black-hole solution of the 5D Einstein equations is given for the metric and two scalar fields. The black-hole 
metric is asymptotically AdS with an event horizon located at a finite value of the bulk coordinate. In the 
5D gravity dual, one scalar field plays the role of a string-theory dilaton that is dual to a gluon operator. The other scalar field resembles the 
string-theory tachyon and is dual to chiral operator.

We attempt to introduce rigor into the soft-wall thermodynamics, beyond what has been done in \cite{Herzog:2006ra, Colangelo:2009ra}. The solution outlined here is an expansion only valid in the limit of $z, z_h<1$. Unfortunately, this is the price we pay to study the soft-wall thermodynamics; the equations of motion have no known closed-form solution nor a reasonable numerical solution. 

The thermodynamics resulting from our solution leads to interesting consequences. We show that a nonzero thermal condensate function $\mathcal{G}$ induces a phase transition \cite{Gursoy:2008za, Gursoy:2008bu, Gubser:2008ny, Galow:2009kw, Megias:2010ku}. Furthermore, $\mathcal{G}$ contributes leading-order terms to the soft-wall thermodynamics that are absent in the lattice results \cite{Boyd:1996bx, Miller:2006hr, Panero:2009tv}. However, qualitatively the two agree. The entropy has the expected behavior at high temperatures, scaling as $T^3$. The speed of sound through the thermal plasma is consistent with the conformal value of $1/3$, matching the upper bound advocated in \cite{Cherman:2009tw}. 

Our analysis begins in Section \ref{secDynamicAct}, where we present the thermal AdS and black-hole AdS solution and compute the on-shell action. In Section \ref{secThermo}, we study the thermodynamics of our solution, including a general expression for the entropy and squared speed of sound. We then calculate the free energy difference by carefully matching our two solutions at the AdS boundary. This enables us to compute the transition temperature between the confined and deconfined phases of the gauge theory in Section \ref{secCond}. Concluding remarks are given in Section \ref{secDiscuss}.

\section{Finite-Temperature Action} 
\label{secDynamicAct}

We begin by specifying the 5D action in the string frame and show how it is related to the 5D action in the Einstein frame. Because the equations of motion become simpler, we transform the action to the Einstein frame. The action is then converted to a finite-temperature description by introducing a black hole into the metric. We generalize the AdS solution with two scalar fields~\cite{Batell:2008zm} to a thermal AdS solution, which is used to compute the 
on-shell action. The black-hole solution will be given in Section~\ref{secbhAdS}.

\subsection{5D Lagrangian}
We start with the string-frame action inspired by the dimensionally-reduced type IIB supergravity action \cite{Son:2007vk, Batell:2008zm},
\begin{eqnarray} 
\label{equString}
\mathcal{S}_{\rm string} &=&-\frac{1}{16 \pi G_5} \int d^5 x \sqrt{-g} \Bigg[{\rm e}^{-2\Phi} 
\Bigg(R + 4\,g^{MN}\partial_M\Phi\partial_N\Phi - \frac{1}{2}\,g^{MN}\partial_M \chi\partial_N \chi 
- \nonumber\\ 
&&\qquad\qquad\qquad\qquad V_S(\Phi, \chi)\Bigg)+\, {\rm e}^{-\Phi}\mathcal{L}_{{\rm meson}}\Bigg] + S_{GH}^{(s)}.
\end{eqnarray}
where $\mathcal{L}_{{\rm meson}}$ contains all the mass terms not considered in this work, $V_S$ is the string-frame scalar potential, and the indices $M,N = (t,x_{1},x_{2},x_{3},z)$. We work with two scalar fields $\Phi$ and $\chi$ where previous models have included only the dilaton \cite{BallonBayona:2007vp, Gursoy:2008za, Gursoy:2008bu, Alanen:2009na, Alanen:2010tg, Gubser:2008ny, Franco:2009if, Galow:2009kw, Megias:2010ku}. The inclusion of a Gibbons-Hawking term, $\mathcal{S}_{GH}^{(s)}$, is an attempt to be more rigorous than previous models. The string-frame metric is assumed to have an AdS-Schwarzschild form
\begin{equation} \label{equStringMetric}
ds^2 = g_{MN} dx^M dx^N= \frac{R^{2}}{z^{2}}\left(-f(z) dt^{2} + d\vec{x}^{2} + \frac{dz^{2}}{f(z)}\right), 
\end{equation}
where $f(z)$ determines the location of the black-hole horizon. Furthermore, the string-frame action (\ref{equString}) contains similarities with noncritical string theory. The $\Phi$ scalar field behaves like a dilaton, while the $\chi$ scalar field behaves like a closed-string tachyon field. This suggests that our setup could be the low-energy limit of some underlying string theory.

While the string-frame action provides a suitable starting point, it is more practical to do calculations in the Einstein frame. Switching to the Einstein frame involves a simple conformal transformation,
\begin{equation} \label{equconformal}
g_{MN}^{(s)} = {\rm e}^{\frac{4}{3}\Phi}g_{MN}^{(E)}.
\end{equation}
The gravity-dilaton-tachyon action in the Einstein frame then becomes
\begin{equation} 
\mathcal{S}_E =-\frac{1}{16 \pi G_{5}} \int d^5x \sqrt{-g}\left(R - \frac{1}{2}(\partial\phi)^2 - \frac{1}{2}(\partial\chi)^2- V(\phi, \chi)\right)
+ S_{GH},
\label{equEinstein}
\end{equation}
where $\phi = \sqrt{8/3} \Phi$ and $V = V_S\,{\rm e}^{\frac{4}{3}\Phi}$. We explicitly define the Gibbons-Hawking term $S_{GH}$ as
\begin{equation}
\mathcal{S}_{GH} = \frac{1}{8 \pi G_5}\int d^4 x \sqrt{-\gamma} K , \label{GHaction}
\end{equation}
where $K=\gamma^{\mu\nu}K_{\mu\nu}$ and $\gamma$ is the four-dimensional induced metric at the AdS boundary. The extrinsic curvature $K_{\mu\nu}$ is defined by
\begin{equation}\label{defnintcurv}
      K_{\mu\nu} = \frac{1}{2} n^M \partial_M \gamma_{\mu\nu}~,
\end{equation}
 where the vector $n_{M}$ is the outward directed normal to the boundary and
\begin{equation}
g_{MN}\,n^{M}n^{N} = 1.
\end{equation}
The boundary of the AdS$_{5}$ space considered here is the $z=0$ plane, making the normal vector 
\begin{equation}
      n^M = -\frac{1}{\sqrt{g_{zz}}} \left(\frac{\partial}{\partial z}\right)^M =  \frac{\delta^M_z}{\sqrt{g_{zz}}}.
      \label{defnormal}
\end{equation}
The Gibbons-Hawking term does not affect the equations of motion, but will have consequences when considering the free energy and deconfinement temperature.

To introduce temperature, we compactify the Euclidean time coordinate, 
$\tau\equiv i t\rightarrow it+ \beta$, where $\beta=1/T$ is the inverse temperature. The 
finite-temperature metric in the Einstein frame then becomes
\begin{equation} 
\label{equEinsteinMetric}
ds^{2}=a(z)^{2}\left(f(z)d\tau^2+d\vec{x}^{2}+\frac{dz^{2}}{f(z)}\right),
\end{equation}
with the finite-temperature action given by
\begin{eqnarray}
\mathcal{S}_E(\beta)&=&-\frac{1}{16 \pi G_{5}} \int d^4 x \int_0^{\beta} d\tau \int dz \sqrt{-g}\left(R - \frac{1}{2}(\partial\phi)^2 - \frac{1}{2}(\partial\chi)^2 - V(\phi, \chi)
\right)\nonumber \\
&&\qquad\qquad +\frac{1}{8 \pi G_5}\int d^3 x \int_0^{\beta} d\tau \sqrt{-\gamma} K.
\label{equEinstein1}
\end{eqnarray}
The Einstein frame metric (\ref{equEinsteinMetric}) and action (\ref{equEinstein1}) will be 
used to solve the Einstein's equations in two realms, thermal AdS (thAdS) and black-hole AdS (bhAdS). All quantities with a subscript $_{0}$ are associated with the thAdS solution.

We fix the value of the 5D gravitational coupling $G_{5}$ using the QCD matching found in \cite{Gursoy:2008za}, making
\begin{equation}
G_{5} = \frac{45\pi R^{3}}{16 N_{c}^{2}}.
\end{equation}

\subsection{Thermal AdS Solution}
\label{secThermalAdS}

We begin by obtaining a thAdS solution that is effectively equivalent to the zero-temperature case explored in \cite{Batell:2008zm}, where $f(z)=1$.  Assuming the scalar fields are a function of only the $z$ coordinate, the action can be expressed as
\begin{eqnarray} 
\label{equAction1}
\mathcal{S}_0(\delta) &=& -\frac{N_c^2}{45\pi^2}\frac{V}{R^3 T} \int_{\delta}^{\infty}{dz \sqrt{-g}\left(R-\frac{1}{2}g^{55}\phi'^{2} - \frac{1}{2}g^{55}\chi'^{2} - V(\phi,\chi)\right)}\nonumber\\
 && \qquad\qquad\qquad+S_{0,GH},
\end{eqnarray}
where $V$ is the spatial three-volume. A UV cutoff at $z=\delta$ has been introduced to regularize 
any singular behavior at the AdS boundary. 

 The metric associated with the thAdS solution is
\begin{equation} 
\label{equEmetric}
      ds^2 = a(z)^{2}(d\tau^2 + d\vec{x}^2 + dz^2) \equiv {\rm e}^{-2 c\,\phi(z)}\frac{R^2}{z^2}(d\tau^2 + d\vec{x}^2 + dz^2),
\end{equation}
where the constant $c$ depends on the conformal transformation. In our case, $c=1/\sqrt{6}$.

 Two equations of motion come from the 5D Einsteins equations,
\begin{eqnarray}
12 \frac{a'^{2}}{a^{2}} - 6 \frac{a''}{a} &=& \phi'^{2} + \chi'^{2}, \label{equGeneralField} \\
6\frac{a'^{2}}{a^{2}} + 3 \frac{a''}{a} &=& -a^{2} V(z), \label{equGeneralV}
\end{eqnarray}
and two more equations come from the varying of the action,
\begin{eqnarray}
a^{2}\frac{\partial V}{\partial \phi} &=& \phi'' + 3 \phi' \frac{a'}{a}, \label{equVphi} \\ 
a^{2}\frac{\partial V}{\partial \chi} &=& \chi'' + 3 \chi' \frac{a'}{a}, \label{equVchi}
\end{eqnarray}
where prime $(')$ denotes derivatives with respect to $z$. 
Using (\ref{equEmetric}), we express (\ref{equGeneralField}), (\ref{equGeneralV}), (\ref{equVphi}), and (\ref{equVchi}) all in terms of $\phi$, $\chi$, and $V\left(\phi(z),\chi(z)\right)$,
\begin{eqnarray}
\chi'^{2} &=& \frac{2\sqrt{6}}{z}\phi' + \sqrt{6} \phi'', \label{equGeneralField2}\\
V(z) &=& \frac{{\rm e}^{\frac{2}{\sqrt{6}} \phi}}{R^2}\left(-12 - 3\sqrt{6} z \phi' - \frac{3z^{2}}{2} \phi'^{2} + \sqrt{\frac{3}{2}} z^{2} \phi''\right), \label{equGeneralV2}\\
\frac{\partial V}{\partial \phi} &=& \frac{z^{2}{\rm e}^{\frac{2}{\sqrt{6}}\phi}}{R^{2}} \left(\phi'' - \sqrt{\frac{3}{2}} \phi'^{2} - \frac{3}{z} \phi'\right),\label{equVphi2}\\
\frac{\partial V}{\partial \chi} &=& \frac{z^{2}{\rm e}^{\frac{2}{\sqrt{6}}\phi}}{R^{2}} \left(\chi'' - \sqrt{\frac{3}{2}} \phi'\,\chi' - \frac{3}{z} \chi'\right),\label{equVchi2}
\end{eqnarray}
where we see that the nonlinear term $\phi'^{2}$ has conveniently cancelled in (\ref{equGeneralField2}). Of the four equations, we find that three are independent.

As shown in \cite{Batell:2008zm}, the solution in the soft-wall model with a quadratic dilaton gives
\begin{eqnarray}
\phi(z) &=& \sqrt{\frac{8}{3}} \mu^2 z^2, \label{equphisol}\\
\chi(z) &=& 2 \sqrt{6} \mu z, \label{equchisol}\\
V(z) &=& \frac{{\rm e}^{\frac{4}{3} \mu^2 z^2}}{R^2} \left(-12 - 20 \mu^2 z^2 - 16 \mu^4 z^4\right), \label{equpotsol}
\end{eqnarray}
where $\mu$ sets the hadronic mass scale. The quadratic behavior of the $\phi$ solution (\ref{equphisol}) leads to a Regge-like hadron mass spectrum $(m_n^2\sim \mu^2 n)$. The potential is a function of the $z$ coordinate with no unique solution for $V(\phi,\chi)$. However, a potential was found in \cite{Batell:2008zm},
\begin{equation}\label{equbatellV}
V(\phi,\chi) = \frac{\chi}{2}{\rm e}^{\frac{\chi^{2}}{18}} + 2 \phi^{2}{\rm e}^{\frac{2}{\sqrt{6}} \phi} - 12 \left[3 {\rm e}^{\frac{\chi}{36}}-2\left(1-\frac{2}{\sqrt{6}}\right){\rm e}^{\frac{\phi}{\sqrt{6}}}\right]^{2}.
\end{equation}
We can define an alternative potential that still produces (\ref{equphisol}), (\ref{equchisol}), and (\ref{equpotsol}),
\begin{equation}\label{equkelleyV}
V_{\rm alt}(\phi,\chi) = \frac{{\rm e}^{\frac{2}{\sqrt{6}}\phi}}{R^{2}}\left(-12 + 4\sqrt{6} \phi - \frac{3}{2}\chi^{2} - 4\phi^{2} + \frac{7}{3\sqrt{6}}\phi\chi^{2}-\frac{2}{27}\chi^{4}\right).
\end{equation}
More examples of consistent and well-defined potentials written in terms of two scalar fields can be found in \cite{Kapusta:2010mf}.

Part of the the free energy expression is found by substituting the thAdS solution into (\ref{equAction1}) and finding the Gibbons-Hawking term. In general, we find the Ricci scalar and extrinsic curvature, 
\begin{eqnarray} \label{equRicciS1}
R &=& -\frac{8 a''}{a^{3}} - \frac{4 a'^{2}}{a^{4}},\\
 \gamma^{\mu\nu} K_{\mu\nu} &\equiv& K = 4\frac{a_0'}{a_0^2}.
\end{eqnarray}
Substituting into (\ref{GHaction}) gives the on-shell Gibbons-Hawking term for the
thermal AdS solution,
\begin{equation} \label{equGHterm0}
\mathcal{S}_{0,GH} = \frac{N_c^2V}{45\pi^2} \frac{8 a_0^2 a_0'}{R^3 T} .
\end{equation}
the on-shell action (\ref{equAction1}) then becomes
\begin{equation}
\mathcal{S}_0(\delta) = \frac{N_c^2\,V}{15\pi^2} \frac{2 a'(\delta) a^2(\delta)}{R^3 T}.
\label{equS1}
\end{equation}
The on-shell action is a pure boundary term and strictly depends on the AdS boundary conditions as $\delta\rightarrow 0$.

\subsubsection{Field/Operator Correspondence}\label{secFOcorr}

According to the AdS/CFT dictionary, a dimension-$\Delta$ operator in $d$-dimensional gauge theory corresponds to a scalar field in the gravity dual with a mass,
\begin{equation}
m^{2} = \Delta(\Delta - d).
\end{equation} 
Expanding the potential (\ref{equbatellV}) from \cite{Batell:2008zm}, we see that 
\begin{equation}\label{equAcase}
m_{\phi}^{2}R^{2} = -4, \quad\quad\quad m_{\chi}^{2}R^{2} = -3,
\end{equation}
suggesting that $\phi$ is dual to a dimension-2 operator, and $\chi$ is dual to a dimension-3 operator. However, the potential (\ref{equkelleyV}) gives the masses as
\begin{equation}\label{equGcase}
m_{\phi}^{2}R^{2} = 0, \quad\quad\quad m_{\chi}^{2}R^{2} = -3,
\end{equation}
indicating that $\phi$ and $\chi$ are dual to a dimension-4 and dimension-3 operator, respectively. It is fairly clear that the chiral operator, $q\bar{q}$, and the gluonic operator, Tr$\left[F^{2}\right]$, are the dimension-3 and dimension-4 operators, but the dimension-2 operator is much less clear. 

The most likely dimension-2 operator candidate is $A_{\mu}^{2}$, which becomes a local expression in the Laudau gauge, $\partial^{\mu}A_{\mu}=0$. Coupling a source term to $A_{\mu}^{2}$ makes the theory nonrenormalizable at the quantum level. A quadratic source term can be added to remedy this obstacle, though, this ruins the energy interpretation of the effective action \cite{Vercauteren:2010rk}. In the context of the AdS/CFT correspondence, $A_{\mu}^{2}$ is often understood to convey information about the topological defects in the gravity dual \cite{Gubarev:2000eu, Gubarev:2000nz}. Much more work concerning $A_{\mu}^{2}$ has been conducted in \cite{Dudal:2009tq, Vercauteren:2010cg, Vercauteren:2011ze}.

In the current soft-wall case, the field/operator correspondence appears complicated. The ambiguity stems from the fact that the original AdS/CFT dictionary was formulated considering purely free scalar fields. The potentials (\ref{equbatellV}) and (\ref{equkelleyV}) clearly have interaction terms. Resolving the issue of whether interaction terms affect the field/operator correspondence and determining the interpretation of the dimension-2 operator is the subject of future research. In that spirit, we expand upon the published potential (\ref{equbatellV}) and assume that the fields $\chi$ and $\phi$ correspond to the operators $q\bar{q}$ and Tr$\left[F^{2}\right]$. The significance of our work relies on the fact that the dilaton is dual to \emph{some} temperature-dependent operator. The identity of that operator is a topic for further research.

\subsection{Black-Hole AdS solution}
\label{secbhAdS}

Next, we consider the black-hole AdS solution that describes a deconfined phase, mimicking a free quark-gluon plasma. 
Assuming the solutions are only a function of the $z$ coordinate, the 5D action associated with the black-hole AdS solution simplifies to
\begin{eqnarray} 
\label{equAction2}
\mathcal{S}_{bh}(\delta) &=& -\frac{N_{c}^{2}}{45 \pi^{2}}\frac{V}{R^3 T(z_{h})} \int_{\delta}^{z_h}{dz \sqrt{-g}\left(R-\frac{1}{2}g^{55}\phi'^{2} - \frac{1}{2}g^{55}\chi'^{2} - V(\phi,\chi)\right)} \nonumber\\
&&\qquad\qquad\qquad\qquad+S_{bh,GH},
\end{eqnarray}
where $z_h$ is the location of the black-hole horizon. 
We will see that $z_h$ is directly related to the temperature of the gauge theory. We begin with the black-hole metric (\ref{equEinsteinMetric}) and find four independent equations of motion,
\begin{eqnarray}
f''(z) &=& -3 f'(z)\frac{a'(z)}{a(z)}, \label{equbkf}\\
\phi'(z)^{2} + \chi'(z)^{2} &=& 12 \frac{a'(z)^{2}}{a(z)^{2}} - 6 \frac{a''(z)}{a(z)},\label{equbkfields}\\
a(z)^{2}\frac{\partial V}{\partial \phi} &=& f(z)\phi''(z) + f'(z)\phi'(z) + 3 f(z)\phi'\frac{a'(z)}{a(z)}, \label{equbkVphi}\\
a(z)^{2}\frac{\partial V}{\partial \chi} &=& f(z)\chi''(z) + f'(z)\chi'(z) + 3 f(z)\chi'\frac{a'(z)}{a(z)}. \label{equbkVchi}
\end{eqnarray}
Unlike in the thAdS case, the potential is already determined. We must use the potential (\ref{equbatellV}) to connect the soft-wall action to the free energy investigated in Section \ref{secThermo}. With four independent equations and four unknown functions, $f(z)$, $a(z)$, $\phi(z)$, and $\chi(z)$, we cannot assume a fixed relation between the warp factor $a(z)$ and the dilaton $\phi(z)$. These quantities must evolve independently as the temperature varies. The system of equations associated with bhAdS are difficult to solve but for the simplest cases. 

Using the series expansions, we construct another solution. We use the thAdS solution of Section \ref{secThermalAdS} as the starting points for these series expansions,
\begin{eqnarray}
a(z) &=& \frac{R}{z}{\rm e}^{-\frac{\phi}{\sqrt{6}} + \sum_{n=2}^{\infty} m_{n}(\mathcal{G},z_h) z^{n}}, \label{equseriesmetric} \\
\phi(z) &=& \sqrt{\frac{8}{3}}\mu^{2}z^{2} + \sum_{n=2}^{\infty} p_{n}({\cal G},z_h) z^n, \label{equseriesphi} \\
\chi(z) &=& \sum_{n=1}^{\infty} c_{n}(\mathcal{G},z_h) z^{n},\label{equserieschi}\\
f(z) &=& 1 + \sum_{n=4}^{\infty} f_{n}(\mathcal{G},z_{h}) z^{n},\label{equseriesf}
\end{eqnarray}
where we see that one of the black-hole conditions, $f(0)=1$, is automatically satisfied.
The condensate function $\mathcal{G}(z_h)$ plays an important role in the free energy and phase transition of the system.
By solving (\ref{equbkf}), (\ref{equbkfields}), (\ref{equbkVphi}), and (\ref{equbkVchi}) in successive powers of $z$, we find the coefficients up to $n=8$ in terms of $f_{4}(\mathcal{G},z_h)$. The other black-hole condition, $f(z_h)=0$, determines the final unknown coefficient. We calculate the non-zero coefficients for the metric,
\begin{eqnarray}
m_{2}  &=& -\frac{\sqrt{6}}{4\mu^{2}}\mathcal{G}(z_h),\\
m_{6}  &=& \frac{1}{8\mu^{4} + 3\sqrt{6}\mathcal{G}(z_h)}\Bigg( -\frac{8\mu^{6}}{21}f_{4} + \frac{3\sqrt{3}\mu^{2}}{7\sqrt{2}}f_{4}\mathcal{G}(z_h) + \frac{32\sqrt{2}\mu^{6}}{21\sqrt{3}}\mathcal{G}(z_h) \nonumber\\
&& \quad+ \frac{45}{56\mu^{2}}f_{4}\mathcal{G}(z_h)^{2} + \frac{131\mu^{2}}{42}\mathcal{G}(z_h)^{2} +\frac{9\sqrt{3}}{8\sqrt{2}\mu^{2}}\mathcal{G}(z_h)^{3}\nonumber\\
&&\quad - \frac{93}{112\mu^{6}}\mathcal{G}(z_h)^{4} \Bigg),\\ 
m_{8}  &=& \frac{1}{8\mu^{4}+3\sqrt{6}\mathcal{G}(z_h)}\Bigg(-\frac{4\mu^{8}}{9}f_{4} + \frac{11\mu^{4}}{21\sqrt{6}}f_{4}\mathcal{G}(z_h) +\frac{184\sqrt{2}\mu^{8}}{243\sqrt{3}}\mathcal{G}(z_h) \nonumber\\
&&\quad + \frac{71}{56}f_{4}\mathcal{G}(z_h)^{2} + \frac{2543\mu^{4}}{1701}\mathcal{G}(z_h)^{2} + \frac{1507}{378\sqrt{6}}\mathcal{G}^{3} +\frac{117\sqrt{3}}{224\sqrt{2}\mu^{4}}f_{4}\mathcal{G}(z_h)^{3}\nonumber\\
&&\quad + \frac{203}{432\mu^{4}}\mathcal{G}(z_h)^{4} - \frac{895}{896\sqrt{6}\mu^{8}}\mathcal{G}(z_h)^{5} \Bigg),
\end{eqnarray}
the field $\phi$,
\begin{eqnarray}
p_{4}  &=& -\mathcal{G}(z_h),\\ 
p_{6}  &=& -\frac{\mu^{2}}{\sqrt{6}}f_{4} - \frac{\mu^{2}}{3}\mathcal{G}(z_h) + \frac{11}{8\sqrt{6}\mu^{2}}\mathcal{G}(z_h)^{2},\\
p_{8}  &=& \frac{1}{8\mu^{4} + 3\sqrt{6}\mathcal{G}}\Bigg(-\frac{80\sqrt{2}\mu^{8}}{21\sqrt{3}}f_{4} - \frac{12\mu^{4}}{7}f_{4}\mathcal{G}(z_h) - \frac{1408\mu^{8}}{567}\mathcal{G}(z_h) \nonumber\\
&&\quad + \frac{3\sqrt{6}}{7}f_{4}\mathcal{G}(z_h)^{2} -\frac{107\sqrt{2}\mu^{4}}{63\sqrt{3}}\mathcal{G}(z_h)^{2}-\frac{29}{27}\mathcal{G}(z_h)^{3}\nonumber\\
&&\quad-\frac{3083}{1008\sqrt{6}\mu^{4}}\mathcal{G}(z_h)^{4}\Bigg),
\end{eqnarray}
the field $\chi$,
\begin{eqnarray}
c_{1}  &=& \sqrt{24\mu^{2}+\frac{9\sqrt{6}}{\mu^{2}}\mathcal{G}(z_h)},\\
c_{3}  &=& \frac{\frac{\sqrt{3}}{2\mu^{4}}\mathcal{G}(z_h)^{2}-2\sqrt{2}\mathcal{G}(z_h)}{\sqrt{8\mu^{2}+\frac{3\sqrt{6}}{\mu^{2}}\mathcal{G}(z_h)}},\\
c_{5}  &=& \frac{1}{\sqrt{8\mu^{4}+3\sqrt{6}\mathcal{G}(z_h)}}\Bigg(-\sqrt{3}\mu^{3}f_{4} - \frac{9\sqrt{2}}{8\mu}f_{4}\mathcal{G}(z_h) - 3\sqrt{2}\mu^{3}\mathcal{G}(z_h)\nonumber\\
&&\quad - \frac{5\sqrt{3}}{4\mu}\mathcal{G}(z_h)^{2} + \frac{9\sqrt{2}}{8\mu^{5}}\mathcal{G}(z_h)\Bigg),\\
c_{7}  &=& \frac{1}{\sqrt{8\mu^{4}+3\sqrt{6}\mathcal{G}(z_h)}}\Bigg(-\frac{20\mu^{5}}{7\sqrt{3}}f_{4} - \frac{39\mu}{14\sqrt{2}}f_{4}\mathcal{G}(z_h) - \frac{568\sqrt{2}\mu^{5}}{189}\mathcal{G}(z_h)\nonumber\\
&&\quad - \frac{111\sqrt{3}}{112\mu^{3}}f_{4}\mathcal{G}(z_{4})^{2} - \frac{1123\mu}{126\sqrt{3}}\mathcal{G}(z_h)^{2} - \frac{67}{24\sqrt{2}\mu^{3}}\mathcal{G}(z_h)^{3} \nonumber\\
&&\quad+ \frac{151\sqrt{3}}{224\mu^{7}}\mathcal{G}(z_h)^{4} \Bigg),
\end{eqnarray}
and $f(z)$,
\begin{eqnarray}
f_{4} &=& \frac{-1}{z_{h}^{4}\left(1+\frac{4\mu^{2}}{3}z_{h}^{2}+\mu^{4}z_{h}^{4} + \left(\frac{\sqrt{6}}{2\mu^{2}} + \frac{\sqrt{6}}{2}\right)z_{h}^{2}\mathcal{G}(z_h) + \frac{27}{32\mu^{4}}z_{h}^{4}\mathcal{G}(z_h)^{2}\right)},\\
f_{6} &=& \frac{4\mu^{2}}{3}f_{4} + \frac{3}{\sqrt{6}\mu^{2}}f_{4}\mathcal{G}(z_h), \\
f_{8} &=& \mu^{4} f_{4} + \frac{\sqrt{6}}{2}f_{4}\mathcal{G}(z_h) + \frac{27}{32\mu^{4}}f_{4}\mathcal{G}(z_h)^{2}.
\end{eqnarray}

In general, the coefficients of the $z^n$ terms in $p_{n}$ and $m_{n}$ have direct consequences for the free energy. For coefficients $0<n<4$, the free energy contains divergences of the power $n$, while for $n>4$ the function ${\cal G}(z_h)$ does not affect the free energy. Only the $n=4$ coefficients affect the free energy expression. As we will see in Section~\ref{secThermo}, the condensate function ${\cal G}$ then plays a crucial role in giving rise to a finite transition temperature.

An expression for the on-shell action can be obtained by simplifying (\ref{equAction2}). The induced metric in this case is
\begin{equation}\label{equinducedbh}
\gamma = a(z)^{2}\left( f(z)d\tau^{2} + d\vec{x}^{2} \right);
\end{equation} 
therefore, the Gibbons-Hawking action term becomes
\begin{equation}\label{equSbhgh}
\mathcal{S}_{bh,GH} = \frac{N_{c}V}{45\pi^{2}R^{3}T}\left(8 f \,a^{2}a' + f' \,a^{3}\right).
\end{equation} 
Given that the Ricci tensor is 
\begin{equation}
R = -\frac{4 f a'^{2}}{a^{4}} - \frac{8 f a''}{a^{3}} - \frac{8 f'a'}{a^{3}} - \frac{f''}{a^{2}},
\end{equation}
the total on-shell action can then be written as
\begin{equation}
\mathcal{S}_{bh}(\delta) = \frac{N_c^2}{45\pi^2}\frac{V}{R^3 T} \left(6 f(\delta)a'(\delta) a^2(\delta) +f'(\delta) a^3(\delta) \right).
\label{bhonshell}
\end{equation}
Again, the on-shell action is a pure boundary term and depends only on the AdS boundary conditions.
This action will be used to compute the free energy in Section~\ref{secThermo}.

\subsubsection{Field/Operator Correspondence at Finite Temperature}

The scalar field solutions (\ref{equseriesphi}) and (\ref{equserieschi}) correspond to turning on thermal condensates in the gauge theory. To show the relation between ${\cal G}(z_h)$ and the operator condensates, we assume the kinetic term of a canonically normalized bulk scalar field fluctuation, 
denoted $\omega$, to be 
\begin{equation} 
\label{equgenbulk}
\mathcal{S}_{{\rm bulk}} = \frac{1}{2 L^3}\int d^5 x \sqrt{-g}\, \partial_M \omega\partial^M \omega.
\end{equation}
Let us first consider the scalar field $\phi$, where the coupling to the four-dimensional boundary operator is
\begin{equation} \label{equgenboundary}
\mathcal{S}_{{\rm boundary}} = \int{d^{4}x \,\omega_{\phi}\, {\rm Tr}(F^{2})}.
\end{equation}
In terms of the 't Hooft coupling $\lambda = {\rm e}^{\Phi}$, the Yang-Mills field strength is 
\begin{equation}\label{equYMcouple}
\mathcal{S}_{{\rm boundary}} = -\int{d^{4}x\, \frac{1}{4\lambda} {\rm Tr} (F^{2}) },
\end{equation}
making the dilaton fluctuation, 
\begin{equation}\label{equfluxphi}
\delta \mathcal{S}_{{\rm boundary}}  = \frac{1}{4} \int{ d^4x\, \delta\Phi \,{\rm e}^{-\Phi} {\rm Tr}(F^{2})}.
\end{equation}
We are only interested in the fluctuation $\delta\Phi = \Phi - \Phi_{0}$, which allows one to compute the difference between thermal and vacuum values of $\langle {\rm Tr}F^{2} \rangle$ \cite{Gursoy:2008za, Megias:2010ku}. From (\ref{equseriesphi}), we find that $\delta\Phi=-\sqrt{3/8}{\cal G}(z_h) z^4$. Recall that the relation between the expectation value of a $\Delta$-dimensional operator in $d$-dimensional space and the field is
\begin{equation}
\frac{\omega}{z^{\Delta}} \rightarrow \frac{\langle \mathcal{O} \rangle}{2\Delta - d}.
\end{equation}
The function ${\cal G}(z_h)$ then relates directly to the gluon condensate,
\begin{equation}
\langle {\rm Tr} (F^{a})^{2} \rangle -  \langle {\rm Tr} (F^{a})^{2} \rangle_{0} = 
-\sqrt{\frac{2}{3}} \left(\frac{32 N_c^2\lambda}{45\pi^2}\right) {\mathcal G}(z_h). 
\end{equation} 

A similar correspondence can be obtained for the scalar field $\chi$, where it is tempting to
relate $\chi$ to the three-dimensional operator $q \bar{q}$. In this case, the coefficient of the 
$z^3$ term in (\ref{equserieschi}) is proportional to the renormalized chiral condensate, $\langle q \bar{q}\rangle -\langle q \bar{q}\rangle_{0} $. 
For this correspondence to hold, $\chi$ would need to be a bifundamental field in the gauge theory. This can be done by 
promoting $\chi$ to a bifundamental field $\chi^{ab}$, where $a,b$ are group indices and writing, for example
\begin{equation}
\langle \chi^{ab} \rangle = \chi(z) 
 \left( 
 \begin{array}{cc} 
 1 & 0 \\
 0 & 1
\end{array}\right),
\label{bifundveveqn}
\end{equation}
for an $SU(2)\times SU(2)$ symmetry. 

Under this assumption, we can follow the same procedure as for the $\phi$ field, and derive an operator correspondence for $\chi$. Assuming a bulk kinetic term of the form (\ref{equgenbulk}) and boundary operator coupling
\begin{equation}
S_{boundary} = \int{d^{4}x \,\omega_{\chi}\,\bar{q}q},
\end{equation}
we find that
\begin{equation}
\langle \bar{q} q\rangle - \langle \bar{q} q \rangle_{0} =
-\frac{2 N_{c}^{2}}{27\pi^{2}}\left(\frac{\mathcal{G}(z_h)}{\mu}-\frac{5}{8}\frac{\mathcal{G}(z_h)^{2}}{\mu^{5}}\right).
\end{equation}
Thus, we see that the function ${\cal G}(z_h)$ is intimately related to the thermal condensates 
$\langle \bar{q}q \rangle$ and $\langle {\rm Tr} F^2 \rangle$. In fact, the ratio of the gluon condensate and the
chiral condensate is given roughly by
\begin{equation}
        \frac{\langle {\rm Tr} (F^{a})^{2} \rangle -  \langle {\rm Tr} (F^{a})^{2} \rangle_{0}}
        {\langle \bar{q} q\rangle - \langle \bar{q} q \rangle_{0}} \approx \frac{16}{5} \sqrt{6} \lambda \mu,
\end{equation}
and is consistent with that obtained in perturbation theory to first order \cite{Shifman:1978bx}.

\section{Thermodynamics} 
\label{secThermo} 

Having determined the black-hole solution of the 5D gravity-dilaton-tachyon system we now
investigate the thermodynamics. We begin with Hawking's black-hole thermodynamics, where the temperature is found,
\begin{eqnarray}
T(z_h) &=& -\frac{\partial_{z}f(z)}{4\pi}\Big|_{z=z_h} = \frac{1}{4\pi \mathcal{P}(z_h)\,a(z_h)^{3}}\nonumber\\
&\approx& \frac{2}{\pi z_{h}}\left(1-\frac{1}{\frac{1}{2} - 3\mu^{2}z_{h}^{2} - \frac{9\sqrt{3}z_{h}^{2}}{4\sqrt{2}\mu^{2}} + \frac{6\mu^{2}(\sqrt{6}z_{h}^{4}\mathcal{G}(z_h)-5)}{12\mu^{2} + 8\mu^{4}z_{h}^{2} + 3\sqrt{6}z_{h}^{2}\mathcal{G}(z_h)} }\right). \label{equexplicitT}
\end{eqnarray}
The quantity $\mathcal{P}$ is defined as
\begin{equation}
\mathcal{P}(z) = \int_{0}^{z} dx \,a(x)^{-3}.
\end{equation}
Of course, all thermodynamic quantities will depend on the condensate function, whose behavior we address in Section \ref{secCond}.

\subsection{Entropy} \label{subsecEntropy}
The entropy is found in the usual way from black-hole thermodynamics, a subject extensively covered in \cite{Bekenstein:1973,Bekenstein:1974,Hawking:1974sw}. The entropy $S$ is defined as 
\begin{equation}
S = \frac{A_{bh}}{4 G_{5}}.
\end{equation}
As expected for a relativistic gas at high temperature, the entropy density $s(T)$ behaves as $T^3$ at high temperature. 
In our model, we can also calculate the subleading temperature behavior and check that it is consistent.
We compute the entropy density from the area of the black-hole horizon using the induced metric $\gamma$. Using the metric (\ref{equseriesmetric}), we obtain
\begin{equation} \label{equBHarea}
A_{bh} = \int d^3x \,\sqrt{\gamma} = \frac{V\,R^3}{z_h^3}{\rm e}^{-3\left( \frac{\phi}{\sqrt{6}} + m2\,z_{h}^{2} + m6\,z_{h}^{6} + m8\,z_{h}^{8}\right)} ,
\end{equation}
where $V$ is the spatial volume. Using the first few terms in our metric expansion, the entropy density is then given by
\begin{equation}
s = \frac{4N_{c}^{2}V}{45\pi}\left(\frac{1}{z_{h}^{3}} - \frac{2\mu^{2}}{z_h} - \frac{3\sqrt{3}\mathcal{G}}{2\sqrt{2}\mu^{2}z_h} + 2\mu^{4}z_{h} + 2\sqrt{6} z_{h} \mathcal{G} + \frac{27 z_{h}\mathcal{G}^{2}}{16\mu^{4}}\right).
\end{equation}

\subsection{Speed of Sound} \label{secvsound}
With the temperature and entropy known, we also determine the speed of sound through the deconfined medium of the gauge theory. The speed of sound $v_s$ characterizes the hydrodynamic evolution of the deconfined, strongly coupled plasma. It has been suggested that $v_s^2$ in holographic models obeys an upper limit of 1/3 \cite{Cherman:2009tw}.
This can be checked for our solution using the relation
\begin{eqnarray} 
v_s^2 &=& \frac{s \frac{dT}{dz_h}}{T \frac{ds}{dz_h}} = \frac{d\log{T}}{d\log s} \nonumber\\
&=& -1 - \frac{1}{3}\frac{d\log{\mathcal{P}}}{d\log{a}}.
\end{eqnarray}
In our model, the exact form of $v_{s}^{2}$ greatly depends on the form of $\mathcal{G}$. However, we do confirm that 
\begin{equation} 
\frac{d\log{\mathcal{P}}}{d\log{a}} \rightarrow -4,
\end{equation}
in the high temperature limit, recovering the conformal limit with an upper bound of $v_s^2=1/3$.

\subsection{Free Energy} \label{secFEnergy}

The free energy of the deconfined phase is calculated in two ways. We use thermodynamic identities to define the free energy as
\begin{equation}
\mathcal{F} = -\int{S\, dT}= \mathcal{F}_{{\rm min}}-\int{s\, V \frac{dT}{dz_{h}}dz_{h}}. \label{equfreeEdefine}
\end{equation}
where we have found the Bekenstein entropy and have an expression for $T(z_h)$. We have explicitly written the integration constant $\mathcal{F}_{{\rm min}}$ since the problem with using (\ref{equfreeEdefine}) is common among energy definitions, setting the zero point. Because $z_h(T)$ is a multi-valued function, the integral is non-trivial. Calculating the value $\mathcal{F}_{{\rm min}}$ generally requires the expression of entropy in the large-$z_h$ region, exactly where our expanded solutions are invalid. While the integral in (\ref{equfreeEdefine}) is completely calculable, we find the free energy by using the on-shell action. Finding the zero-point energy using actions is much easier; one merely subtracts the background free energy as defined by the thAdS action,
\begin{equation}\label{equfreeEaction}
\mathcal{F} = {\rm lim}_{\delta\rightarrow 0} T\left[\mathcal{S}_{bh}(\delta) - \mathcal{S}_0(\delta)\right].
\end{equation}
Both $\mathcal{S}_{bh}$ and $\mathcal{S}_{0}$ have been computed earlier. 

Before evaluating (\ref{equfreeEaction}), we must properly match the thermal AdS and black-hole AdS metrics at the boundary, $\delta\rightarrow 0$ \cite{Witten:1998zw}. This requires matching the intrinsic geometry of the two solutions at the boundary cut-off \cite{Gursoy:2008za}, where
\begin{eqnarray}
a_0(\delta) &=& a(\delta)\sqrt{f(\delta)}, \nonumber\\
V_{0}a_0(\delta)^{3} &=& V a(\delta)^{3}.\label{equmatch}
\end{eqnarray}
In order for (\ref{equmatch}) to be satisfied, we must evaluate the thAdS and bhAdS solutions at different cut-off points, $\tilde{\delta}$ and $\delta$ respectively. We find that 
\begin{equation}
\tilde{\delta} = \frac{\sqrt{8\mu^{2}\delta^{2} - 2\sqrt{6}\mathcal{G}\delta^{4} + \frac{3\sqrt{6}\mathcal{G}}{\mu^{2}}\delta^{2}}}{2\sqrt{2}\mu}.
\end{equation}
Combining the matching with (\ref{equfreeEaction}), we obtain a rather simple expression for the free energy,
\begin{equation}\label{equfreeEmod2}
\mathcal{F} = \frac{N_c^2 V}{45\pi^2 R^3} {\rm lim}_{\delta\rightarrow 0} \left(6 f(\delta) a(\delta)^{2}a'(\delta) + f' a(\delta)^{3} -6 \sqrt{f(\delta)} a(\delta)^4 \frac{a_0'(\tilde{\delta})}{a_0(\tilde{\delta})^2}\right),
\end{equation}
which can be reduced to
\begin{eqnarray} \label{equfreeEfG}
\mathcal{F} = \frac{N_c^{2}V}{45\pi^2}\left(f_{4}(z_h,\mathcal{G}) + 2\sqrt{6} \mathcal{G} \right) = \frac{2N_c^{2}V}{45\sqrt{6}\pi^2}\mathcal{G} -\frac{1}{4} T S.
\end{eqnarray}
It should be noted that we also checked that the black-hole energy $E$ satisfies the thermodynamic formula 
$E={\cal F} + T S$ by computing the ADM energy in Appendix \ref{appADM}.

\section{The Condensate Function and the Phase Transition}\label{secCond}

All the thermodynamic relationships rely on the behavior of the condensate function; therefore, we need to find a solution to $\mathcal{G}$ to evaluate the temperature, entropy, and speed of sound. The free energy of the system gives us enough information to solve for $\mathcal{G}$. We only need to set (\ref{equfreeEdefine}) and (\ref{equfreeEfG}) equal to one another. Taking the derivative with respect to $z_h$ removes any unknown constants, giving
\begin{equation}\label{equdiffforG}
\frac{df_{4}}{d z_h} + 2\sqrt{6}\frac{d\mathcal{G}}{d z_h} = s\,V\,\frac{dT}{d z_h},
\end{equation}
or in more simplified terms,
\begin{equation} \label{equdiffG}
    \frac{d\mathcal{G}(z_h)}{dz_h} = \frac{1}{2 \sqrt{6} \mathcal{P}(z_h)} \left(\frac{a'(z_h)}{a(z_h)} + \frac{\mathcal{P}'(z_h)}{4 \mathcal{P}(z_h)}\right).
\end{equation}
Unfortunately, (\ref{equdiffG}) is a stiff equation. One can often find stable solutions to these equations within a certain region, but our case is barely within a region of stability. 

\begin{figure}[h!]
\begin{center}
\includegraphics[scale=0.45]{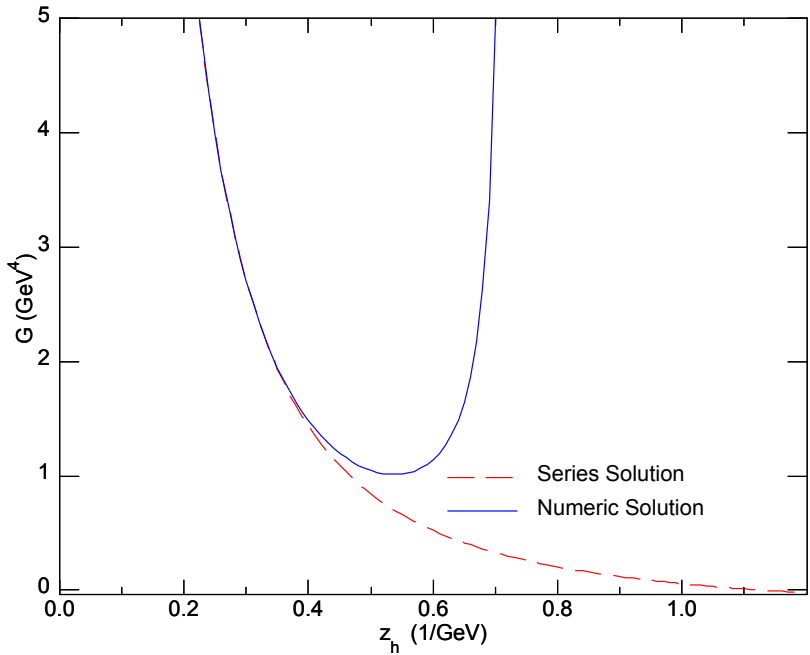}
\caption{The series and numerical solution for $\mathcal{G}$ is plotted. We use the series solution for the thermodynamics since the differential equation clearly gives a numerically unstable solution for $z\ge 0.3$ GeV$^{-1}$.}
\label{figcondensate}
\end{center}
\end{figure}

 We use two methods to solve for $\mathcal{G}$. First, we introduce a series expansion for $\mathcal{G}$,
\begin{equation}\label{equGseries}
\mathcal{G}(z_h) = \sum_{j=-\infty}^{\infty} g_{j} \mu^{j+4} z_{h}^{j},
\end{equation}
and expand (\ref{equdiffG}) in powers of $z_h$. Performing the expansion and matching coefficients, we find that the useful nonzero $g_{j}$'s are
\begin{eqnarray}\label{equGcoeff}
g_{-2}\equiv g &=& 1.43290,\nonumber\\
g_{0} &=& -6.09417.\nonumber\\
\end{eqnarray}
Since the nonlinear nature of (\ref{equdiffG}) spoils the series solution quite quickly, we find that only the first two terms give an accurate solution for $G$ in the range of $z_h<1$. Using the series expansion to inform the boundary conditions, we then numerically solve for $\mathcal{G}$ in (\ref{equdiffG}). However, it is valid within a limited range of $z_{h}$, matches the series solution in that range, and diverges around $z\sim 0.5$ GeV$^{-1}$. The two solutions are plotted in Figure \ref{figcondensate}. With the condensate function $\mathcal{G}$, we can take a second look at the thermodynamics.

Given the expression (\ref{equseriesf}), the temperature can be written as 
\begin{eqnarray}
T(z_h) &\approx& \frac{2}{\pi z_h}\left(\frac{16+12\sqrt{6}g + 27g^{2}}{32+12\sqrt{6}g + 27g^{2}}\right) + \frac{\mu^{2}z_{h}^{2}}{\pi(32+12\sqrt{6}g + 27g^{2})^{2}}\times\nonumber\\
&&\qquad\Bigg(\frac{2048}{3}+516\sqrt{6}g+192 g^{2}+256\sqrt{6}g_{0} + 1728 g g_{0}\nonumber\\
&&\qquad+\, 216\sqrt{6} g^{2} g_{0}\Bigg).\label{equTexactwithG}
\end{eqnarray}
We clearly see that the condensate function contributes a finite piece to the temperature expression. Increasing the strength of the condensate produces 
\begin{equation}\label{equTlimit}
\lim_{g\rightarrow \infty}\,T(z_h)\rightarrow \frac{2}{\pi z_h},
\end{equation}
which is a factor of 2 larger than the conformal limit found in \cite{Gursoy:2008za, Gursoy:2009kk}. We find that other work in this area has assumed that the condensate terms are suppressed logarithmically. However, in the construction of this model, the soft-wall set-up has no natural means of generating the logarithmic suppression in $\mathcal{G}$. As a result, we see that the condensate contributes to leading-order behavior in the small-$z_h$/large-$T$ limit.

It will be useful to have an inverted function $z_{h}(T)$ which we use to transform functions of $z_h$ to functions of $T$. For simplicity, we use only the first-order term,
\begin{equation}\label{equTexpand}
z_{h}(T) \approx \frac{2}{\pi\,T}\left(\frac{16+12\sqrt{6}g + 27g^{2}}{32+12\sqrt{6}g + 27g^{2}}\right).
\end{equation} 

\begin{figure}[h!]
\begin{center}
\includegraphics[scale=0.45]{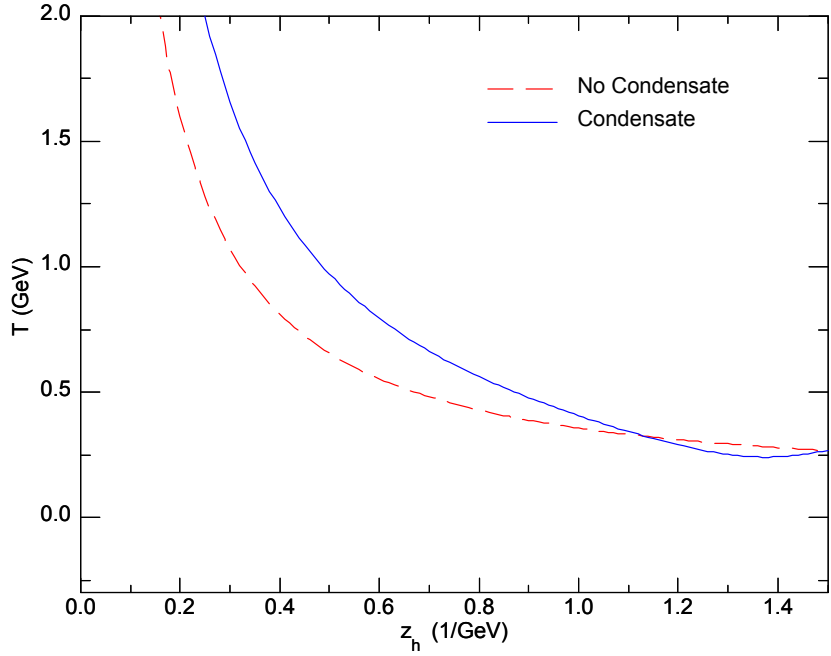}
\caption{The temperature as a function of the horizon location $z_h$ with and without a condensate 
$\mathcal{G}$. The inclusion of a condensate in this particular model precludes a second, unstable black-hole 
solution from developing.}
\label{figTemp}
\end{center}
\end{figure}

\begin{figure}[h!]
\begin{center}
\includegraphics[scale=0.45]{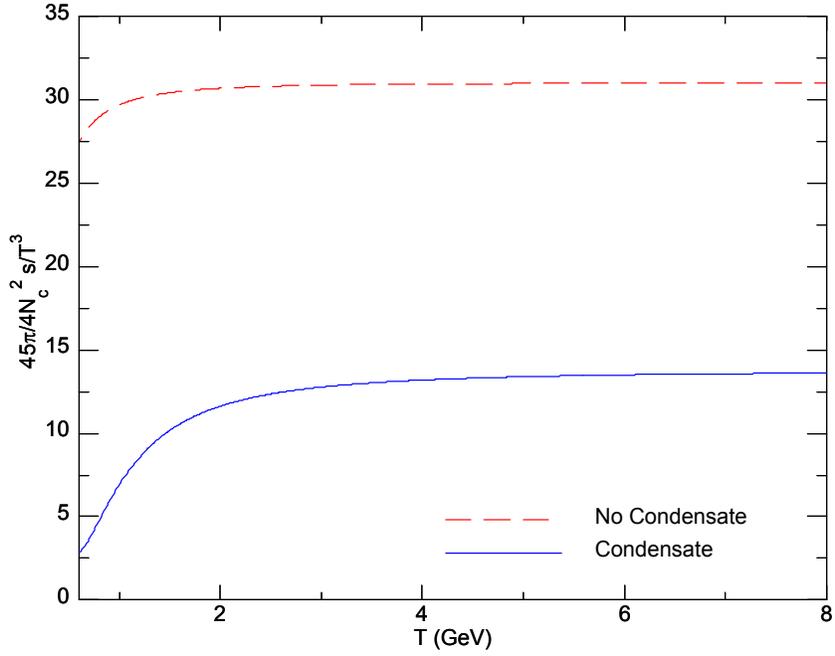}
\caption{The entropy density divided by $T^3$ as a function of temperature. At high temperatures the entropy density slowly evolves to the conformal case of $s\sim T^3$ in each case. }
\label{figEntropy}
\end{center}
\end{figure}

Analytically, the entropy of the deconfined phase can be written in terms of $z_h$, 
\begin{eqnarray}
s(z_h) &=& \frac{4N_c^2}{45\pi}\frac{1}{(32+16\sqrt{6}g+27g^{2})^{2}}\Bigg[32 -\frac{86\sqrt{6}}{7}g +\frac{279}{14}g^{2} - \frac{313\sqrt{3}}{14\sqrt{2}} \nonumber\\
&&\quad + \frac{3623}{168}g^{4} - \frac{313}{224\sqrt{6}}g^{5} + \frac{90081}{3584}g^{6}\Bigg] +\ldots\nonumber\\
&\approx& \frac{1.77}{z_{h}^{3}} - \frac{25.35\mu^{2}}{z_{h}} + 156.27 \mu^{4} z_{h} +\ldots,\label{equspecificentropy} 
\end{eqnarray}
or in terms of $T$,  
\begin{equation}
s(T) \approx 13.88 T^{3} - 50.35 \mu^{2}\,T +\ldots = C(T) T^{3},\label{equGEntropy}
\end{equation}
where the function $\mathcal{C}(T)$ modifies the entropy behavior at low temperatures.
Our result, plotted in Figure \ref{figEntropy}, agrees qualitatively with the lattice data presented in \cite{Boyd:1996bx}. The high-temperature limit of $s(T)/T^3$, however, is shifted from lattice results because of the contribution from the condensate function.

\begin{figure}[h!]
\begin{center}
\includegraphics[scale=0.45]{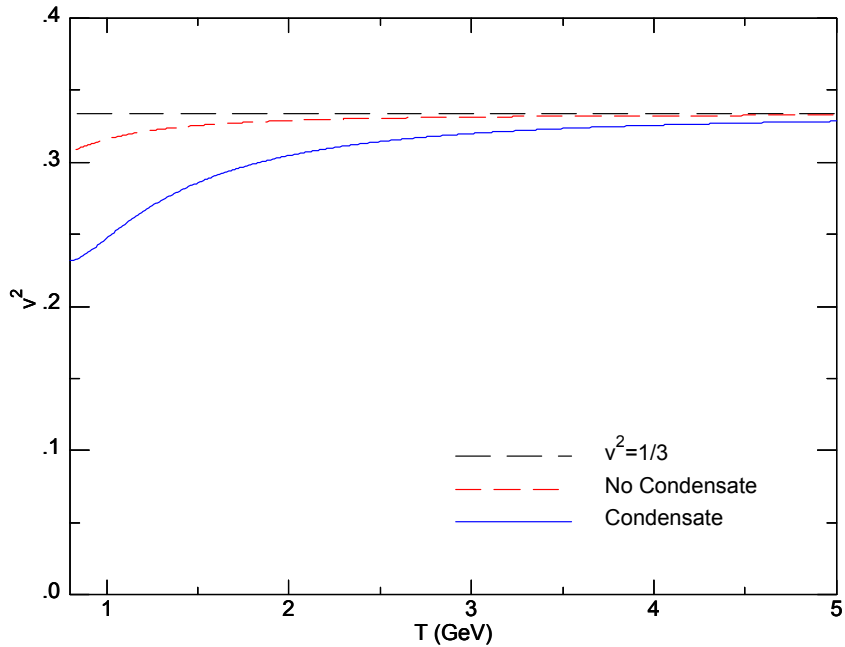}
\caption{The squared speed of sound in the strongly coupled plasma as a function of the temperature. The conformal limit of $v_s^2 = \frac{1}{3}$ is recovered at high temperatures.}
\label{figVelo}
\end{center}
\end{figure}

The speed of sound through the QGP-like thermal phase can be expressed as 
\begin{eqnarray}
v_{s}^{2}(z_h) &\approx& \frac{1}{3} - 2.718\mu^{2}z_{h}^{2} + 21.938 \mu^{4}z_{h}^{4},\label{equv2zh}\\
v_{s}^{2}(T)   &\approx& \frac{1}{3} - 0.689\frac{\mu^{2}}{T^{2}} + 1.409\frac{\mu^{4}}{T^{4}}.\label{equv2T}
\end{eqnarray} 
We clearly see that (\ref{equv2zh}) and (\ref{equv2T}) give the expected $v_{s}^{2} = 1/3$ in the small-$z$ and large-$T$ limit. The speed of sound through our QGP is plotted in Figure \ref{figVelo}. As the figure shows, $v_{s}^{2}$ deceases more rapidly with the condensate terms included; however, the difference between the two cases are not as stark as it was in the entropy case.

An issue arises when we consider the behavior of $\mathcal{G}$. In Section \ref{secFOcorr}, we argued that the dual operator corresponding to the dilaton field is either a dimension-2 or a dimension-4 gluon operator. On dimensional grounds, previous work asserts that $\mathcal{G}\sim T^{4}$. However, our model clearly shows that any self-consistent solution in the soft-wall model requires that $\mathcal{G}\sim  T^{2}$. The consequences of the temperature dependence on the field/operator duality are subjects for further research. 

\subsection{Deconfinement Temperature} 
\label{secDeTemp}

Finding the deconfinement temperature requires examining the free energy of the deconfined phase in (\ref{equfreeEfG}). The point at which $\mathcal{F}=0$ occurs when the black-hole solution becomes more energetically favored than the thermal solution. It is instructive to first consider the case without a condensate. By setting ${\cal G}(z_h)=0$ in (\ref{equfreeEfG}), the transition temperature is determined from the condition
\begin{equation}
\label{equHerzog}
f_{4}(\mathcal{G}(z_h),z_h)=0.
\end{equation} 
However, (\ref{equHerzog}) has only one possible solution, $z_{h}\rightarrow\infty$; therefore, no transition temperature exists since $z_h\rightarrow\infty$ occurs in the region of the unstable black hole. This mimics the scenario considered in \cite{Herzog:2006ra, Colangelo:2009ra}, but we include the full back-reaction of the scalar field and take the Gibbons-Hawking term into account. Therefore, we see that a nonzero condensate is needed to obtain a transition temperature in our model. 

\begin{figure}[h!]
\begin{center}
\includegraphics[scale=0.45]{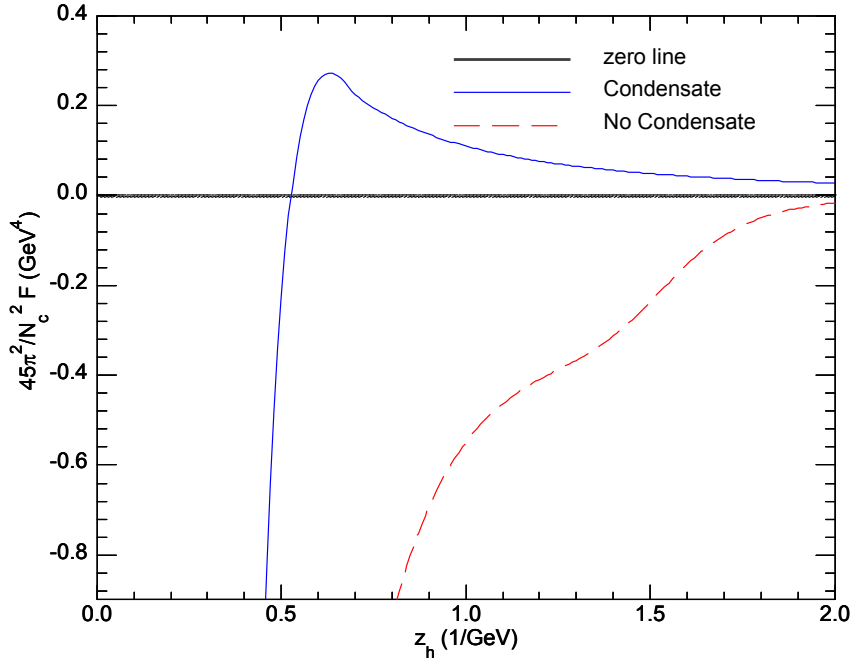}
\caption{The free energy $\cal F$ plotted as a function of $z_h$. Only when including a gluon condensate do we find a solution to $\mathcal{F} = 0$. The free energy with no gluon condensate approaches zero as $z_h\rightarrow\infty$, but never crosses the x-axis.}
\label{figEnergy}
\end{center}
\end{figure}

Using the leading behavior and the numerical solution, we find the free energy behavior and plot it in Figure \ref{figEnergy}. The phase transition occurs at a critical $z_c=0.5262$. Using (\ref{equexplicitT}) and the first two terms of the expansion for $\mathcal{G}$, this corresponds to a critical temperature of $T_{c}=919$ MeV. This is much larger than either theoretical reasoning or lattice calculations suggest \cite{Shifman:1978bx,Shifman:1978bw,Shifman:1978by, Boyd:1996bx, Miller:2006hr}. There are no current plans (that this author can find) for experiments that would produce such high temperatures; however, experimental evidence for QGP has already been detected. The large transition temperature is most likely the product of the crude soft-wall model and the absence of any suppression terms in $\mathcal{G}$. As $z_h$ increases ($T$ decreases), we see that free energy goes to zero as in \cite{Gursoy:2008za, Megias:2010ku}; however, the free energy expression is valid below the the region of $z_h\approx 1$. Thus, the true behavior of the free energy at increasing values of $z_h$ requires more numerical work. 

\section{Conclusion}  
\label{secDiscuss}

We have presented a five-dimensional, gravity-dilaton-tachyon, black-hole, dynamical solution that represents a dual description of a strongly coupled gauge theory at finite temperature. The solution generalizes the soft-wall geometry considered in \cite{Batell:2008zm}, which generates a quadratic dilaton and Regge mass trajectory ($m_n\sim\sqrt{n}$). In the string frame, the dilaton and tachyon fields appear to be duals to the gluon and chiral operators, respectively. However, the actual field/operator correspondence is ambiguous. The black-hole solution describes a deconfined free-gluon phase. A transition from confined matter exists, predicted at a extremely high $T_{c}=919$ MeV. A condensate function is needed for any transition to occur.

 Despite being a mere series expansion, we argue that it gives insight into thermodynamics of a strongly coupled gauge theory produced by the soft-wall model. We expected to find two major contributions to the thermodynamics: the underlying conformal limit and condensate terms. Although we considered a finite number of terms in our infinite solution expansion, our calculations suggest that the condensate terms produce convergent quantities, as shown in (\ref{equTlimit}). With further study and numerical rigor, we believe a viable closed-form, black-hole solution with a lower transition temperature will be found.

While our model has some features reminiscent of QCD at finite temperature, it still represents a crude approximation with a number of shortcomings. The bhAdS solution is only valid in the region of small $z_h$. The power-law dependence of the scalar fields and metric does not include the logarithmic corrections needed to suppress condensate terms as in \cite{Gursoy:2008za, Megias:2010ku}, resulting in noticeable changes in the behaviors of the temperature, entropy, and speed of sound. Furthermore, using the AdS/CFT dictionary, the dilaton field appears to be dual to a dimension-2 operator, $(A_{\mu})^{2}$, in contradiction to the standard assumption of a dimension-4 operator, $Tr[F^{2}]$. Nevertheless, our 5D dynamical black-hole solution with two scalar fields provides a toy model to understand the nontrivial properties of strong-coupled gauge theories at finite temperature.

\section*{Acknowledgements} 
\label{acknowl}

I like to thank Tony Gherghetta for his contributions and advice. I also thank Joe Kapusta for his continuing support and Christopher Herzog for his clarification on calculating deconfinement temperature. I would like to thank Francesco Nitti for a critical reading of an earlier draft 
version of this manuscript. The work was supported by the US Department of Energy (DOE) under Grant No. DE-FG02-87ER40328 and by the Doctoral Dissertation Fellowship from the Graduate School at University of Minnesota.

\vspace{40pt}
\appendix

\section*{APPENDIX}
\section{The Arnowitt-Deser-Misner (ADM) Energy}
\label{appADM}
The ADM energy is a useful definition of energy for gravitational systems approaching an asymptotic, well-defined metric at the boundary \cite{Arnowitt:1961zz,Hawking:1995fd}. To verify the thermodynamic relation $E={\cal F}+T\,S$, we will compute the ADM energy for our
black-hole solution with respect to the thermal AdS solution. Considering a time slicing of the 5D metric in ADM form,
\begin{equation}
     ds^2 = -N^2 dt^2 +\gamma_{ij}(dx^i-N^i dt)(dx^j-N^j dt),
\end{equation}
with $i,j=x_{1},x_{2},x_{3},z$. In our case, we have
\begin{eqnarray}
N=a(z)\sqrt{f(z)},\\
\gamma = a(z)^{2}\left(d\vec{x}^{2} + \frac{dz^{2}}{f(z)}\right).
\end{eqnarray}

The  expression for the ADM energy is given by \cite{Gursoy:2008za, Hawking:1995fd},
\begin{equation}
      E=-\frac{1}{8\pi G_5} \int d\Sigma_B N\left( \sqrt{-\gamma_{\rm ind}} ^{(3)}K -  \sqrt{-\gamma_{0,\rm{ind}}} ^{(3)}K_0 \right)~.
\end{equation}
The integral is performed over a three-dimensional surface at the bulk boundary $\Sigma_B$ embedded in the 4D constant time slice $\Sigma_t$, where $\gamma_{\rm ind}$ is the induced three-dimensional metric. The superscript $^{(3)}$ refers to the dimensionality of the extrinsic curvature. Using three-dimensional equivalents of (\ref{defnintcurv}) and (\ref{defnormal}), we obtain
\begin{equation}
       E=\frac{2 N_c^2}{15\pi^2}\frac{V}{L^3} a^2(\delta) \sqrt{f(\delta)}\left( \sqrt{f(\delta)}a'(\delta) - a^2(\delta)
       \frac{a_0'(\delta)}{a_0^2(\delta)}\right),
\end{equation}
where we have matched the two solutions at the AdS boundary as specified in \ref{equmatch}. Taking the limit $\delta\rightarrow 0$ gives the 
final expression
\begin{equation}
       E=-\frac{N_c^2V}{15\pi^2} \left(f_{4}(z_h,\mathcal{G}) -\frac{2 \sqrt{6}}{3} {\cal G}(z_h)\right) = {\cal F} + T S,
\end{equation}
where we have used the fact that
\begin{equation}
T\,S = -\frac{4 N_{c}^{2} V}{45 \pi^{2}} f_{4}. 
\end{equation}

\bibliographystyle{h-physrev4}
\bibliography{finiteRef}

\end{document}